\documentclass[a4paper,11pt]{article}
\pdfoutput=1 
\usepackage{jcappub, bm, color} 
\usepackage{amssymb,amsfonts,slashed,amsthm,amsmath,graphicx,soul,empheq}
\usepackage[caption=false]{subfig}
\usepackage{ulem}
\begin{document}

\newcommand{\vev}[1]{ \left\langle {#1} \right\rangle }
\newcommand{\bra}[1]{ \langle {#1} | }
\newcommand{\ket}[1]{ | {#1} \rangle }
\newcommand{\eV}{ \ {\rm eV} }
\newcommand{\KeV}{ \ {\rm keV} }
\newcommand{\MeV}{\  {\rm MeV} }
\newcommand{\GeV}{\  {\rm GeV} }
\newcommand{\TeV}{\  {\rm TeV} }
\newcommand{\1}{\mbox{1}\hspace{-0.25em}\mbox{l}}
\newcommand{\Red}[1]{{\color{red} {#1}}}

\newcommand{\lmk}{\left(}  
\newcommand{\rmk}{\right)}
\newcommand{\lkk}{\left[}  
\newcommand{\rkk}{\right]}
\newcommand{\lhk}{\left \{ }  
\newcommand{\rhk}{\right \} }
\newcommand{\del}{\partial}  
\newcommand{\la}{\left\langle} 
\newcommand{\ra}{\right\rangle}
\newcommand{\half}{\frac{1}{2}}

\newcommand{\bea}{\begin{array}}
\newcommand{\eea}{\end{array}}
\newcommand{\beq}{\begin{eqnarray}}
\newcommand{\eeq}{\end{eqnarray}}
\newcommand{\eq}[1]{Eq.~(\ref{#1})}

\newcommand{\dd}{\mathrm{d}}
\newcommand{\Mpl}{M_{\rm Pl}}
\newcommand{\mg}{m_{3/2}}
\newcommand{\abs}[1]{\left\vert {#1} \right\vert}
\newcommand{\mphi}{m_{\phi}}
\newcommand{\Hz}{\ {\rm Hz}}
\newcommand{\for}{\quad \text{for }}
\newcommand{\Min}{\text{Min}}
\newcommand{\Max}{\text{Max}}
\newcommand{\Kahler}{K\"{a}hler }
\newcommand{\cphi}{\varphi}
\newcommand{\Tr}{\text{Tr}}
\newcommand{\diag}{{\rm diag}}

\newcommand{\SUf}{SU(3)_{\rm f}}
\newcommand{\Upq}{U(1)_{\rm PQ}}
\newcommand{\Zpq}{Z^{\rm PQ}_3}
\newcommand{\Cpq}{C_{\rm PQ}}
\newcommand{\ubar}{u^c}
\newcommand{\dbar}{d^c}
\newcommand{\ebar}{e^c}
\newcommand{\nubar}{\nu^c}
\newcommand{\Ndw}{N_{\rm DW}}
\newcommand{\Fpq}{F_{\rm PQ}}
\newcommand{\fpq}{v_{\rm PQ}}
\newcommand{\Br}{{\rm Br}}
\newcommand{\Lag}{\mathcal{L}}
\newcommand{\Lqcd}{\Lambda_{\rm QCD}}

\newcommand{\ji}{j_{\rm inf}} 
\newcommand{\jb}{j_{B-L}} 
\newcommand{\M}{M} 
\newcommand{\im}{{\rm Im} }
\newcommand{\re}{{\rm Re} }

\def\lrf#1#2{ \left(\frac{#1}{#2}\right)}
\def\lrfp#1#2#3{ \left(\frac{#1}{#2} \right)^{#3}}
\def\lrp#1#2{\left( #1 \right)^{#2}}
\def\REF#1{Ref.~\cite{#1}}
\def\SEC#1{Sec.~\ref{#1}}
\def\FIG#1{Fig.~\ref{#1}}
\def\EQ#1{Eq.~(\ref{#1})}
\def\EQS#1{Eqs.~(\ref{#1})}
\def\TEV#1{10^{#1}{\rm\,TeV}}
\def\GEV#1{10^{#1}{\rm\,GeV}}
\def\MEV#1{10^{#1}{\rm\,MeV}}
\def\KEV#1{10^{#1}{\rm\,keV}}
\def\blue#1{\textcolor{blue}{#1}}
\def\red#1{\textcolor{blue}{#1}}

\newcommand{\eff}{\Delta N_{\rm eff}}
\newcommand{\neff}{\Delta N_{\rm eff}}
\newcommand{\cc}{\Omega_\Lambda}
\newcommand{\Mpc}{\ {\rm Mpc}}
\newcommand{\Msolar}{M_\odot}

\def\ft#1{\textcolor{blue}{#1}}
\def\FT#1{\textcolor{blue}{[{\bf FT:} #1]}}
\def\sn#1{\textcolor{red}{#1}}
\def\SN#1{\textcolor{red}{[{\bf SN:} #1]}}
\def\WY#1{\textcolor{magenta}{#1}}
\def\WYC#1{\textcolor{magenta}{[{\bf WY:}{\bf #1}]}}

\begin{flushright}
TU-1167
\end{flushright}

\title{
Early dark energy by dark Higgs, and axion-induced non-thermal trapping
}

\author{
Shota Nakagawa,
}
\author{
Fuminobu Takahashi,
}
\author{
Wen Yin
}
\affiliation{Department of Physics, Tohoku University, 
Sendai, Miyagi 980-8578, Japan}

\abstract{
We propose a new scenario of early dark energy (EDE) with a dark Higgs trapped at the origin. To keep this dark Higgs trapped until around the matter-radiation equality,  we use dark photons produced non-thermally by coherent oscillations of axions, which have a much stronger trapping effect than thermal mass. When the trapping ends, the dark Higgs quickly decays into dark photons, which are then red-shifted as radiation. 
 The dark Higgs EDE scenario works well for an ordinary Mexican-hat potential, and the dark Higgs naturally sits at the origin from the beginning, since it is the symmetry-enhanced point. Thus, unlike the axion EDE, there is no need for elaborate potentials or fine-tuning with respect to the initial condition.
 Interestingly, the axion not only produces dark photons, but also explains dark matter. We find the viable parameter region of the axion decay constant  and the axion mass  where dark matter and the $H_0$ tension can be simultaneously explained. 
We also discuss the detectability of the axion in the presence of axion-photon coupling, and show that the axion can be the QCD axion.
}

\emailAdd{shota.nakagawa.r7@dc.tohoku.ac.jp}
\emailAdd{fumi@tohoku.ac.jp}
\emailAdd{yin.wen.b3@tohoku.ac.jp}

\maketitle
\flushbottom

\section{Introduction
\label{introduction}}
The currently observed universe is largely explained by the flat Friedmann-Lema$\hat{\text{\i}}$tre-Robertson-Walker (FLRW) spacetime with the cosmological constant and cold dark matter (CDM), known as 
the standard $\Lambda$CDM paradigm. However, as the accuracy of various observations has improved in recent years, several tensions between different  observational data  have become apparent. The most prominent of these is the tension related to the current Hubble constant, $H_0$, which is called the Hubble tension. There is a significant discrepancy between the Hubble constant measured directly from local observations and the value inferred from measurements of the cosmic microwave background radiation (CMB) under the assumption of the $\Lambda$CDM cosmology. Specifically, the value of $H_0$ inferred from the Planck data of the CMB anisotropies assuming $\Lambda$CDM is given by $H_0=67.4\pm0.5{\rm km/s/Mpc}$ \cite{Planck:2018vyg}. On the other hand, the SH0ES team utilizes the Type-Ia supernovae data and the Hubble Space Telescope observations of Cepheids to obtain $H_0=73.04\pm1.04{\rm km/s/Mpc}$ \cite{Riess:2021jrx}.
The discrepancy between them is at the $5\sigma$ level, if the reported values are taken at face value.
There are a variety of methods for determining $H_0$, e.g. using other calibrators for the cosmic distance ladder methods \cite{Huang:2019yhh,Dhawan:2022yws, Blakeslee:2021rqi}, time-delay strong lensing \cite{Denzel:2020zuq}, etc. See the recent review paper \cite{Perivolaropoulos:2021jda} for an exhaustive summary of these local measurements.

The origin of the Hubble tension is not yet known. It could be due to unknown   systematic errors, but it could also be a sign of unknown physics beyond the $\Lambda$CDM paradigm. Especially in the latter case, some elaborate prescription tends to be needed to raise the value of $H_0$ inferred from CMB, since the angular scale of the sound horizon at the last scattering surface (LSS) has been precisely determined by the CMB observations.
It is given by the ratio of the (comoving) sound horizon $r_s$ to the angular diameter distance $d_A$,
\beq
\theta_s=\frac{r_s}{d_A}\simeq\frac{H_0r_s}{\int_0^{z_{\rm drag}}dz[(1+z)^3\Omega_{\rm m} + (1-\Omega_{\rm m})]^{-1/2}},
\label{angle}
\eeq
where the flat $\Lambda$CDM model is assumed in the second equality, $\Omega_{\rm m}$ is the density parameter for non-relativistic matter, $z_{\rm drag}$ is the redshift when the photon pressure can no longer drag baryons which then become gravitationally unstable, and we neglect the radiation components in the late-time universe. While the integral in the denominator is determined by the late-time cosmology, $r_s$, in the numerator depends on the early-time cosmology as, $a_0r_s\equiv\int^{\infty}_{z_{\rm drag}}dz c_s(z)/H(z)$ where $a_0$ denotes the current scale factor, $c_s$ the sound speed for the baryon-photon fluid, and $H(z)$ the Hubble parameter as a function of the redshift $z$.
Therefore, one basic idea to increase $H_0$ while keeping the angular scale $\theta_s$ the same is to slightly modify the evolution of the early 
 universe to reduce $r_s$ without affecting the late universe.

One of the plausible solutions along this line is the {early dark energy (EDE) \cite{Poulin:2018cxd,Agrawal:2019lmo,Smith:2019ihp,Niedermann:2020dwg}, which behaves as a cosmological constant until some critical redshift $z_c$, typically taken around the matter-radiation equality, and at $z < z_c$, its energy density decreases as radiation or faster. The EDE slightly enhances the expansion rate around $z = z_c$, suppressing the size of the sound horizon. As a result, we would have to get closer to LSS by increasing $H_0$ so as not to alter the corresponding angular scale. In \REF{Poulin:2018cxd}, the authors considered as EDE an axion field $\phi$ with the potential $V(\phi)\propto(1-\cos(\phi/f_\phi))^n$, where $f_\phi$ is the decay constant for the axion and $n$ is a positive integer not less than $2$.\footnote{
The same potential was also studied in the so-called ALP miracle scenario where a single axion field explains both inflation and DM~\cite{Daido:2017tbr,Daido:2017wwb}. See also the multi-natural inflation scenario~\cite{Czerny:2014wza,Czerny:2014qqa} with the same kind of the inflaton potential. 
} Using the dataset including the Planck, BAO measurements, the Pantheon supernovae data, and the SH0ES result of $H_0$, 
they showed that the ratio $f_{\rm EDE}$ of the EDE energy density to the total energy density should be about $0.05$ around $z_c\sim 5000$ in order to relax the $H_0$ tension.

While the EDE scenario with axion is one of the possible solutions to the Hubble tension, there are two points that are non-trivial to satisfy naturally. First, the EDE component must disappear quickly after the critical redshift by the recombination epoch. For this reason,   a rather contrived form of the potential
was assumed. The second difficulty is the fine-tuning regarding the initial value of the EDE scalar field. In order to behave like a cosmological constant around the matter-radiation equality, the initial filed value must be set close enough to the potential maximum, and the fluctuations must also be small enough.  This requires either fine-tuning of the initial condition or non-trivial dynamics such as the mixing with a heavier axion~\cite{Daido:2017tbr,Takahashi:2019pqf,Takahashi:2019qmh}. Also, the question of ``Who ordered that?" remains since the EDE scalar does not play any other cosmological role.

In this paper, we propose a new scenario for the EDE, which does not require any contrived potential nor fine-tuning of the initial condition. In our scenario we use a dark Higgs field trapped at the origin as an EDE component. Such a trapped scalar has been extensively studied in the context of thermal inflation with thermal mass~\cite{Lyth:1995ka,Yamamoto:1985rd}. However,  it is difficult to keep the scalar field trapped by a thermal mass until the matter-radiation equality.\footnote{
Since the effect of trapping by thermal mass is weaker, the dark Higgs needs to be relatively strongly coupled to the lighter particles in thermal plasma. When the dark Higgs settles at the potential minimum, those particles become heavier, and so, other light particles may have to be introduced for successful energy transfer.  In general, the potential of the dark Higgs must be flatter in the case of trapping by thermal mass, and it is non-trivial if the dark Higgs can be EDE in this case.}
Recently, Kitajima and two of the present authors (SN and FT) showed that non-thermally produced dark photons cause far stronger trapping effects,
and that the dark Higgs can be trapped for a very long time and its decay can dilute the cosmologically unwanted relics such as the moduli fields~\cite{Kitajima:2021bjq}.
They considered
tachyonic production of dark photons from coherent oscillations of the axion~\cite{Garretson:1992vt},
which has been studied in a variety of contexts, including the reduction of the QCD axion abundance~\cite{Agrawal:2017eqm, Kitajima:2017peg} and the production of dark photon DM~\cite{Agrawal:2018vin, Co:2018lka,Co:2021rhi}. By using the non-thermal trapping effect, we can trap the dark Higgs for a long time, which plays the role of EDE. After the end of trapping, the dark Higgs promptly decays into dark photons, which will be redshifted as radiation afterwards. The dark Higgs EDE scenario works for an ordinary Mexican-hat potential, and no contrived potential is needed. Also, the dark Higgs naturally sits at the origin from the beginning, since it is the special point in field space where the U(1)$_{\rm H}$ gauge symmetry is restored. 
Interestingly, the axion can naturally explain DM. In particular, we find the viable parameter region where DM and the $H_0$ tension can be simultaneously explained.
As we will see, some mild hierarchy between the axion-dark photon coupling and the gauge coupling is required for our scenario to work, which can be 
explained in a model where there are two U(1) gauge symmetries with a kinetic mixing.  Finally, we will discuss the detectability of the axion in the presence of the axion-photon coupling, and study if the axion can also be the QCD axion.

The rest of this paper is organized as follows. In section \ref{sec:trapping} we briefly review the non-thermal trapping effect and discuss some required initial conditions for the efficient tachyonic production. In section \ref{sec:EDE} we 
study the cosmological evolution of the dark Higgs, axion, and dark photons, and derive various limits on the model parameters so that the dark Higgs plays the role of EDE to solve the Hubble tension. 
In section \ref{sec:model} we present a UV model with two hidden U(1) gauge symmetries which explain the mild hierarchy between the axion-dark photon coupling strength and the gauge coupling constant, and study if the axion can be identified with the QCD axion. Section \ref{sec:conclusion} is devoted for discussion and conclusions.

\section{Non-thermal trapping mechanism
\label{sec:trapping}}
The key to realizing the EDE scenario with dark Higgs is how to trap the dark Higgs at the origin until the matter-radiation equality. Here we consider a mechanism to  trap the dark Higgs by dark photons non-thermally produced by the axion condensate. Such non-thermal trapping is known to be much stronger than trapping by thermal mass. The potential energy of the dark Higgs can explain the EDE for a certain choice of the parameters. When the non-thermal trapping ends, the dark Higgs develops a nonzero vacuum expectation value (VEV) and decays instantly into dark photons. Therefore, the necessary conditions for a successful
EDE scenario are naturally satisfied.

In the following we briefly review the non-thermal trapping mechanism using an Abelian Higgs model with axion, based on \REF{Kitajima:2021bjq}. We also discuss the initial condition for the dark Higgs before the onset of the tachyonic production of dark photons. The detailed UV model for meeting such initial condition and the evaluation of the period of time between when the axion begins to oscillate and when the backreaction of tachyonic production becomes important will be given in Appendix.

\subsection{Set-up
\label{sec:setup}}
First, let us introduce the Abelian Higgs model with axion coupled to the hidden U$(1)_{\rm H}$ gauge boson. The Lagrangian is given by
\beq
\mathcal{L} = (D_\mu\Psi)^\dag D^\mu\Psi-V_\Psi(\Psi,\Psi^\dag)-\frac{1}{4}X_{\mu\nu}X^{\mu\nu}+\frac{1}{2}\del_\mu\phi\del^\mu\phi-V_\phi(\phi)-\frac{\beta}{4f_\phi}\phi X_{\mu\nu}\tilde{X}^{\mu\nu},
\label{Lagrangian}
\eeq
where $\Psi$ is the dark Higgs field with a unit charge, $X_{\mu\nu}=\del_\mu A_\nu-\del_\nu A_\mu$ is the field strength tensor of the hidden U(1)$_{\rm H}$ gauge field $A_{\mu}$, $\tilde{X}^{\mu\nu}=\epsilon^{\mu\nu\rho\sigma}X_{\rho\sigma}/(2\sqrt{-g})$ is its dual with $g\equiv {\rm{det}}(g_{\mu\nu})$, 
and  $\phi$ is the axion. Here $D_\mu=\del_\mu-i e A_\mu$ is the covariant derivative with $e$ being the gauge coupling, $f_\phi$ is the axion decay constant, and $\beta$ is the axion coupling with gauge bosons. As we will see,  $\beta = {\cal O}(10-100)$ is required for efficient tachyonic production~\cite{Kitajima:2017peg}.
We will refer to the gauge boson $A_\mu$ as dark photon in the following, and denote it interchangeably   by $A_\mu$ or $\gamma'$. 

The potentials of the dark Higgs and the axion are given by
\begin{align}\label{higgspot}
V_\Psi(\Psi,\Psi^\dag)&=\frac{\lambda}{4}\left(|\Psi|^2-v^2\right)^2
=V_0 - m_\Psi^2 |\Psi|^2 + \frac{\lambda}{4} |\Psi|^4,\\
V_\phi(\phi)&=m_\phi^2f_\phi^2\left[1-\cos\left(\frac{\phi}{f_\phi}\right)\right],
\end{align}
where $v$ is the VEV of the dark Higgs, $\lambda$ is the quartic coupling, $m_\phi$ is the axion mass, and we define $V_0 \equiv \lambda v^4/4$ and $m_\Psi^2 \equiv \lambda v^2/2$. For simplicity, we assume that the axion mass $m_\phi$ is constant with time in the following analysis. The case of the QCD axion with a temperature dependent  mass will be discussed in \SEC{sec:model}. Note that, in contrast to the original EDE model~\cite{Poulin:2018cxd}, we need only consider the usual Mexican-hat potential; there is no need to consider a contrived form of the potential. This is due to the strong non-thermal trapping and subsequent rapid decay of dark Higgs into dark photons.

For efficient production of dark photons by the axion condensate, the dark Higgs should stay at the origin
until the  axion begins to oscillate. To this end we will introduce a slight modification to the 
above potential. For more information on this, see the end of this section and Appendix. 
For now, we assume that the dark Higgs field is at the origin until the axion starts oscillating.

\subsection{Tachyonic production of dark photons
\label{sec:tachyonic}}
From the Lagrangian (\ref{Lagrangian}), the equations of motion for dark photon and axion in the flat FLRW universe are given by
\beq
\ddot{A}_i+H\dot{A}_i-\frac{1}{a^2}(\nabla^2A_i-\del_i\del_jA_j)-2e{\rm{Im}}(\Psi^*\del_i\Psi)+2e^2|\Psi|^2A_i\nonumber\\
-\frac{\beta}{f_\phi a}\epsilon_{ijk}\left(\dot{\phi}\del_jA_k-\del_j\phi\dot{A}_k\right)=0,\label{DPeom}\\
\ddot{\phi}+3H\dot{\phi}-\frac{1}{a^2}\nabla^2\phi+\frac{\del V_{\phi}}{\del\phi}+\frac{\beta}{f_\phi a^3}\epsilon_{ijk}\dot{A}_i\del_jA_k=0,
\label{axioneom}\\
\del_i\dot{A}_i-2ea^2{\rm{Im}}(\Psi^*\dot{\Psi})-\frac{\beta}{f_\phi a}\epsilon_{ijk}(\del_i\phi)(\del_jA_k)=0,
\label{const}
\eeq
where $a$ is the scale factor,  $H\equiv \dot{a}/a$ is the Hubble parameter, and we adopt the temporal gauge, $A_0 = 0$. Here the overdot represents the derivative with respect to time, and we denote $\partial_i \partial^i = - a^{-2} \partial_i \partial_i = - a^{-2} \nabla^2$.  The tachyonic production of dark photons is induced by
the last term in (\ref{DPeom}). The last equation (\ref{const}) is the constraint equation on  the longitudinal component.

Let us assume that the axion is initially spatially homogeneous due to the primordial inflation, and that 
the U$(1)_{\rm H}$ symmetry is restored, i.e., $\langle\Psi\rangle=0$.
Then, \EQ{DPeom} is simplified as
\beq \label{eq:Apm}
\ddot{A}_{\bm{k},\pm}+H\dot{A}_{\bm{k},\pm}+\frac{k}{a}\left(\frac{k}{a} \mp \frac{\beta\dot{\phi}}{f_\phi}\right)A_{\bm{k},\pm}=0,
\eeq
where $A_{\bm{k},\pm}$ is the Fourier component of the dark photon field in the circular polarization basis. One can see that either of the two circular polarization modes grows exponentially for $k/a < \beta|\dot\phi|/f_\phi$, depending on the sign of $\dot\phi$~\cite{Garretson:1992vt}.
The tachyonic growth of the dark photon is so efficient that the energy density of the dark photon soon becomes comparable to that of the axion. 
The system then enters a nonlinear regime, and such linear analysis is no longer valid.

The tachyonic production of dark photons during the nonlinear regime has been studied in detail by the numerical lattice simulations~\cite{Kitajima:2017peg,Agrawal:2018vin,Kitajima:2021bjq}. 
Numerical results show that once in the nonlinear regime, tachyonic production becomes inefficient and stops when a significant fraction of the axion energy is transferred to dark photons. 
Let us define the spatially averaged value of the physical gauge field as $({\bm A})_i \equiv \sqrt{\langle A^2_i\rangle}/a$, where 
 $A_\mu$ is the comoving field in the expanding universe.  Then,
the typical field value of dark photons when the system enters the nonlinear regime is given by $|{\bm A}_{\rm nl}|\simeq2f_\phi/\beta$,  which is typically a few orders of magnitude smaller than the axion decay constant.\footnote{This can be intuitively understood by noting that most of the axion oscillation energy $\sim m_\phi^2 f_\phi^2 \theta_*^2$ is transferred to dark photons with typical momentum of $\beta m_\phi \theta_*$. Here $\theta_*$ denotes the initial misalignment angle, and the typical momentum of the dark photon is determined by the instability band of Eq.~(\ref{eq:Apm}). The detailed discussion is given in Appendix \ref{sec:app}.}. Here and in what follows the variables evaluated when entering the non-linear regime are labeled with `nl'. 
After the tachyonic production stops, the amplitude of the dark photon decreases with time as $|\bm{A}|\propto a^{-1}$. Such a large field value is the source of the effective mass of the dark Higgs.

To see the effective mass  of the dark Higgs explicitly, we expand the dark Higgs field as $\Psi=(v + \frac{s}{\sqrt{2}})e^{i\theta_\Psi}$, where $s$ is the radial mode, and $\theta_\Psi$ is the dimensionless Nambu-Goldstone mode. 
Then we can rewrite the kinetic term of $\Psi$ as
\beq
|D_\mu\Psi|^2= \frac{1}{2} (\del_\mu s)^2 + \frac{1}{2} (\del_\mu\theta_\Psi-eA_\mu)^2 s^2+\dots~,
\eeq
for $|\Psi| \ne 0$. 
The second term can be interpreted as the gauge-invariant effective mass for the dark Higgs field
when the dark photon acquires large field values.
In Ref.~\cite{Kitajima:2021bjq} it was shown that 
 the contribution of $\partial \theta_\Psi$ is subdominant and the effective mass squared is well 
 approximated by $e^2 {\bm A}^2$ in their numerical lattice simulations. 
It was also shown  that this effective mass traps the dark Higgs
field around the origin until it becomes smaller than the bare mass $m_\Psi^2$.
If this trapping lasts sufficiently long, the dark Higgs potential energy could play an interesting cosmological role
such as EDE or late-time inflation~\cite{Kitajima:2021bjq}.

When the dark photons get redshifted and the curvature of the dark Higgs potential at the origin becomes negative, i.e. $e|{\bm{A}}_{\rm end}|\simeq m_{\Psi}$, the dark Higgs starts to roll down to the potential minimum. Note that the dark Higgs quickly reaches the true minimum, because the amplitude of dark photons decreases as $|{\bm A}|\propto 1/(e|\Psi|)^{1/2} $ when $e |\Psi|$ becomes larger than the momentum. Thus the positive effective mass of the dark Higgs quickly disappears as it develops a nonzero VEV. 
The ratio of the scale factors from the beginning of the non-linear regime to the end of trapping is given by
\beq
\frac{a_{\rm end}}{a_{\rm nl}} = \frac{|{\bm{A}}_{\rm nl}|}{|{\bm{A}}_{\rm end}|}
\simeq \frac{2 e }{\beta}\frac{f_\phi}{m_\Psi},
\label{endnl}
\eeq
where we use the fact $|\bm{A}|\propto a^{-1}$ during the trapping regime.
This factor determines the endpoint of EDE, i.e., the critical redshift $z_c$. In \SEC{sec:EDE} we will determine the viable range of parameters for which the dark Higgs field behaves as EDE.

\subsection{The initial condition of dark Higgs
\label{sec:initial}}
So far we have assumed that the dark Higgs has a temporary positive mass and stays at the origin until the onset of the axion oscillations. This is because it is advantageous with respect to dark photon production in two ways. First, the mass of  dark photons suppresses the tachyonic production. Suppose the dark Higgs has a nonzero VEV, and the dark photon is massive. Then, if the dark photon mass $m_{\gamma'}$ is greater than the typical momentum of the instability band, tachyonic production will not occur. More precisely speaking,  if $m_{\gamma'}\gtrsim\beta|\dot{\phi}|/2f_\phi\sim\beta m_\phi/2$, then the tachyonic production is suppressed~\cite{Agrawal:2018vin}.
Second, during tachyonic production, the electric and magnetic fields of the U(1)$_{\rm H}$ become very strong, and if the dark Higgs is light at that time, many pairs of dark Higgs bosons are produced from vacuum
via the Schwinger effect \cite{Heisenberg:1936nmg,Schwinger:1951nm}. This will greatly hinder the growth of the electric field.\footnote{A similar problem arises when dark photons have a kinetic mixing with hypercharge.
This is because massive dark photons are coupled to the SM sector through the kinetic mixing. On the other hand, since massless dark photons are decoupled from the SM sector even in the presence of the kinetic mixing, one can introduce the kinetic mixing without hampering the tachyonic production. Related issues were also discussed 
in Ref.~~\cite{Agrawal:2018vin}.} Note that,
once dark photons are produced and their characteristic field values become sufficiently large, the dark Higgs becomes so heavy that the pair production through the Schwinger effect becomes exponentially 
suppressed, and it remains trapped for a long time. So, it is essential to stabilize the dark Higgs at the origin with a sufficiently heavy mass before the axion starts oscillating.

To avoid the problems mentioned above, we slightly extend the set-up to give a temporary positive mass to the dark Higgs field so that it stays at the origin and the U(1)$_{\rm H}$ symmetry remains unbroken until the axion starts to oscillate. Typically, the mass should be heavier than the Hubble parameter at that time to suppress the Schwinger effect~\cite{Kitajima:2021bjq}. The effective potential for the dark Higgs is now given by
\beq
V_\Psi^{(\rm eff)}(\Psi,\Psi^\dag) = V_0 + (\Delta m^2 - m_\Psi^2)|\Psi|^2 + \cdots,
\eeq
where $\Delta m^2$ represents the additional mass term.
It should satisfy $\Delta m^2- m_\Psi^2 > {\cal O}(H^2)$
when the axion begins to oscillate at $m_\phi\simeq H$.

To get a feel for how large $\Delta m^2$ should be,  let us estimate the value of $m_\Psi$ required for the dark Higgs to explain EDE. The potential energy of the dark Higgs should be about $5\%$ of the total energy density around the equality~\cite{Poulin:2018cxd}. Assuming the radiation-dominated universe, the dark Higgs potential energy is approximately given by $\lambda v^4\sim0.1 T_c^4=0.1(1+z_c)^4T_0^4$, and we obtain $v\sim1\eV\cdot\lambda^{-1/4}$,  where $T_c$ and $T_0 \simeq 2.725{\rm K}$ are  the temperatures at $z = z_c$ and at present, respectively.
So, the bare mass is roughly given by $m_\Psi \sim 1\eV\cdot\lambda^{1/4}$.
Thus, the additional mass must be comparable to or heavier than ${\cal O}(1)$ eV when $H \sim m_\phi$.
For a more precise estimate, see \SEC{sec:evolution}.

Now we consider a non-minimal coupling to gravity,
\beq
\mathcal{L}\supset-\xi R|\Psi|^2,
\eeq
where $\xi$ is a numerical coefficient, and $R$ is the Ricci curvature.
 The most natural value of $\xi$ is of ${\cal O}(1)$, and we take
 $\xi > 0$ for our purpose. In the flat FLRW universe, the Ricci scalar curvature is given by $R=6[\ddot{a}/a+(\dot{a}/a)^2]$. 
In the matter-dominated universe, we have $R=3H^2$, and the additional mass is given by $\Delta m^2 = 3 \xi H^2$. In order for the dark Higgs to stay at the origin when the axion starts oscillating, we need $m_\phi\gtrsim m_\Psi/\sqrt{3\xi}\sim 0.1 \lambda^\frac{1}{4}/\sqrt{\xi} \eV$ for $\xi \gtrsim {\cal O}(1)$. In the radiation-dominated universe, the Ricci curvature vanishes at the classical level, but quantum effects generically induce small but non-negligible contributions, and we obtain $\Delta m^2 \sim 0.01 \xi H^2$. In this case $\xi$ must be much larger than unity.  Therefore, there is a lower limit on the axion mass to trap the dark Higgs using the non-minimal coupling.

How to trap the dark Higgs first is actually model-dependent, and the range of possible axion masses can be taken even wider. 
See  Appendix \ref{app:trapping}  for a discussion of the case with a portal coupling to the Standard Model Higgs as a specific example.
As for the initial value of the dark Higgs, once the dark photon is created, it has no effect on the subsequent cosmological evolution. So, to be conservative, we will  treat the axion mass as a free parameter, without
going into details of the UV models in the following sections.

\section{Early dark energy by dark Higgs
\label{sec:EDE}}

In the previous section, we have discussed non-thermal trapping effects on the dark Higgs and established the basic setup needed for our scenario. Here we  apply it to our dark Higgs EDE  scenario to determine which parameter space provides a plausible solution to the Hubble tension. Our EDE scenario is closely related to the dynamics of axion. In particular, the axion not only produces dark photons via tachyonic preheating, but can also be DM, making our EDE scenario attractive from a cosmological perspective. The dark Higgs decays to dark photons at the end of the trapping, which then behave as (self-interacting) dark radiation. In the following, we first discuss the properties of the dark Higgs, axion, and dark photon, respectively, and then present the parameter regions allowed by observational and theoretical constraints.

\subsection{Cosmology of dark Higgs EDE}
\label{sec:evolution}
\subsubsection{Dark Higgs}
To alleviate the Hubble tension, the vacuum energy density of the EDE scalar should be several \% of the total energy density around the equality.
According to Ref. \cite{Poulin:2018cxd}, the best-fit parameters  estimated by the Markov Chain Monte Calro simulation are given by $f_{\rm EDE}\simeq5\%$ at $z_c\simeq5000$. We adopt their best-fit parameters as reference values for a successful EDE scenario, because the cosmological evolution of EDE in our scenario is almost the same as theirs except for oscillatory features at $z<z_c$.

The size of $f_{\rm EDE}$ determines the upper bound on the dark Higgs mass. The potential energy of the dark Higgs at the origin is given by
\beq
V_0=\frac{\lambda v^4}{4} = f_{\rm EDE} \,\rho_{\rm tot}(T_c)\simeq f_{\rm EDE}\cdot\frac{\pi^2}{30}g_{*0}T_c^4,
\label{condition}
\eeq
where  $\rho_{\rm tot}(T_c)$ is the total energy density at $z=z_c$, and the radiation-dominated universe is assumed in the last equality. Here $g_{*0} \simeq 3.363$ denotes the effective relativistic degrees of freedom for energy density at that time. Rewriting the above Eq.~(\ref{condition}) by using the dark Higgs mass at the minimum, $m_s=\sqrt{\lambda}v$, we obtain a relation between the quartic coupling and the mass of dark Higgs at the potential minimum,
\beq
\lambda&=&\left(\frac{m_s}{(1+z_c)T_{0}}\right)^4\frac{30}{4\pi^2g_{*0}f_{\rm EDE}},\nonumber\\
&\simeq& 2.4\left(\frac{1+z_c}{5000}\right)^{-4}\left(\frac{f_{\rm EDE}}{0.05}\right)^{-1}\left(\frac{m_s}{1\,\eV}\right)^4.
\label{quartic}
\eeq
Thus, the dark Higgs EDE requires $m_s\lesssim 1\eV$ for our reference values of $f_{\rm EDE}$ and $z_c$, since $\lambda$ cannot be much larger than unity for perturbativity. 
Note that we need
$m_s > 2m_{\gamma'}$, or equivalently $e^2<\lambda/8$, for the dark Higgs to decay into a pair of dark photons, where $m_{\gamma'}\equiv\sqrt{2}ev$ represents the dark photon mass in the vacuum.

\subsubsection{Axion 
\label{sec:axion}}
The axion evolves according to \EQ{axioneom} under the coupling with dark photons. The axion is produced at the onset of the oscillations, but its abundance is quickly reduced to about $10\%$ due to the explosive production of dark photons~\cite{Kitajima:2017peg}. Thereafter, it will soon start to decrease with the expansion of the universe as non-relativistic matter.

The axion abundance is determined by the mass $m_\phi$, decay constant $f_\phi$, and the initial misalignment angle $\theta_*$. The most natural value of $\theta_*$ is of ${\cal O}(1)$, which is assumed below. Let us estimate the temperature at that time, $T_{\rm osc}$, in two ways. First, the onset of the axion oscillation is determined by the relation between the Hubble parameter and the axion mass, $H_{\rm osc}\sim m_\phi$. It is then given by
\beq
T_{\rm osc}=\left(\frac{90}{\pi^2g_{*}(T_{\rm osc})}\right)^{1/4}\sqrt{\Mpl m_\phi},
\eeq
where $\Mpl \simeq 2.4 \times 10^{18}$\,GeV is the reduced Planck mass.

Secondly, we can also estimate when the oscillation begins by following the cosmological history. Using (\ref{endnl}), we obtain the oscillation temperature,
\beq
T_{\rm osc}&\simeq&\frac{a_{\rm end}}{a_{\rm osc}}T_{c}=e(1+z_c)\frac{2f_\phi}{\beta}\frac{T_0}{m_{\Psi}}\frac{a_{\rm nl}}{a_{\rm osc}},\nonumber\\
&\simeq&4.1\times10^4\GeV\,  \lambda^{-1/4}\left(\frac{e}{10^{-7}}\right) \left(\frac{\beta}{100}\right)^{-1}
\left(\frac{f_{\rm EDE}}{0.05}\right)^{-1/4}
\left(\frac{a_{\rm nl}}{10a_{\rm osc}}\right)
\left(\frac{f_\phi}{10^{12}\GeV}\right),~~~~
\label{osc}
\eeq
where we have substituted (\ref{quartic}) in the last equality,  and  $a_{\rm nl}/a_{\rm osc}$ is estimated to be $\mathcal{O}(10)$ (more precisely, see Appendix \ref{sec:app}). Thus, the oscillation timing is determined by the duration of the non-thermal trapping.

Combining the two estimates of $T_{\rm osc}$, we obtain the axion mass,
\beq
m_\phi\simeq2.4\eV\,\lambda^{-1/2}
\left(\frac{e}{10^{-7}}\right)^2 \left(\frac{\beta}{100}\right)^{-2}
\left(\frac{a_{\rm nl}}{10a_{\rm osc}}\right)^2\left(\frac{f_{\rm EDE}}{0.05}\right)^{-1/2}\left(\frac{f_\phi}{10^{12}\GeV}\right)^2,
\label{axionmass}
\eeq
where we assume $g_*(T_{\rm osc})=106.75$. Note that the axion becomes heavier for a larger decay constant. 
This is because the amount of dark photons produced from the axion becomes larger and the non-thermal trapping lasts longer, so that for the dark Higgs to start oscillating near the matter-radiation equality as EDE, the axion must start to oscillate earlier.

With the above oscillating temperature, we can estimate the axion abundance. When the system enters the non-linear regime, a significant amount of the axion energy is dissipated into dark photons. According to \cite{Kitajima:2017peg}, the axion abundance is approximately reduced to $10\%$ for $\beta = \mathcal{O}(10)$. Thus, the final axion abundance is given by
\beq
\Omega_\phi^{(\rm stable)} h^2& \sim &10^{-1}\cdot m_{\phi}\frac{s_0}{\rho_{\rm crit}h^{-2}}\frac{m_\phi\theta_*^2f_\phi^2/2}{s(T_{\rm osc})},\nonumber\\
&\simeq&0.02\,\theta_*^2\left(\frac{g_{*s}(T_{\rm osc})}{106.75}\right)^{-1/4}\left(\frac{m_\phi}{1\eV}\right)^{1/2}\left(\frac{f_\phi}{10^{12}\GeV}\right)^2,\\
 &\simeq&0.02\,\theta_*^2\lambda^{-1/4}\left(\frac{e}{10^{-7}}\right)\left(\frac{\beta}{100}\right)^{-1}\left(\frac{a_{\rm nl}}{10a_{\rm osc}}\right)\left(\frac{f_{\rm EDE}}{0.05}\right)^{-1/4}\left(\frac{f_\phi}{10^{12}\GeV}\right)^3,~~~
 \label{Omhstable}
\eeq
where we have used (\ref{axionmass}) in the last equality, and also assumed that the axion is stable on cosmological time scales, namely, the lifetime of the axion is much longer than the present age of the universe.

The DM axion with small $f_\phi$ is subject to the bound on the lifetime. In our setup, if kinematically allowed, the axion decays into dark photons through the coupling of the last term in \EQ{Lagrangian}. The decay rate is given by
\beq
\Gamma_\phi(\phi\rightarrow\gamma'\gamma')=\frac{\beta^2}{64\pi}\frac{m_\phi^3}{f_\phi^2},
\eeq
and if this is the dominant decay channel, the axion lifetime is
\beq
\tau_\phi\simeq1.3\times10^{25}\sec\left(\frac{\beta}{100}\right)^{-2}\left(\frac{m_\phi}{1\eV}\right)^{-3}\left(\frac{f_\phi}{10^{12}\GeV}\right)^2.
\eeq
Taking account of the decay, the present abundance of the axion is given by $\Omega_{\phi} = \exp(-t_0/\tau_\phi)\cdot\Omega_\phi^{(\rm stable)}$ with $t_0\simeq 13.8\,{\rm Gyr}$ being the present age of the universe. If we focus on the cosmologically stable axion, i.e. $\tau_\phi\gtrsim t_0$, this places a lower bound on the decay constant for a given $m_\phi$.

The observational impact of DM decaying into dark radiation was studied in detail \cite{Enqvist:2015ara,Enqvist:2019tsa,Alvi:2022aam}, and the most recent analysis gives the lower bound on the lifetime, $\tau_{\rm DM}\gtrsim 246{\rm Gyr}$ (95\% CL)~\cite{Enqvist:2019tsa}. 
The bound is significantly relaxed, if the axion abundance is smaller than 10\% of the total DM abundance. Assuming that the axion constitutes the total DM density, the lower bound on the lifetime can be expressed as
\beq
m_\phi\lesssim0.11{\rm \, keV}\,\theta_*^{-4/7}\left(\frac{\beta}{100}\right)^{-4/7}.
\label{lifetime}
\eeq
%

\subsubsection{Dark photon
\label{sec:DR}}
Dark photons are first produced from axion through tachyonic preheating, and after non-thermal trapping is completed, they are produced by the decay of the dark Higgs. In particular, 
for the EDE scenario to work,  the dark Higgs must decay quickly into dark photons. Since the U$(1)_{\rm H}$ is spontaneously broken, the decay is allowed if $m_s>2m_{\gamma'}$. Here we study the abundance and properties of the dark photons. As we shall see below, dark photons behave as self-interacting dark radiation.

The primordial dark photons produced by the tachyonic instabilities do not play any important role unless $e$ is extraordinarily small. Taking account of the redshift of the field value $|{\bm A}|$ and the momentum, we can estimate the number density  of the dark photons at the end of the non-thermal trapping as 
\beq 
n_{\gamma'}^{\rm (pri)} \sim 
{m_\phi}|{\bm A}_{\rm end}|^2 \left(\frac{a_{\rm nl}}{{a}_{\rm end}}\right),
\eeq 
where we have assumed that most of the primordial dark photons are produced when the dynamics become non-linear.
Since the dark Higgs immediately reaches the potential minimum after the non-thermal trapping ends, the number density remains the same during the symmetry breaking.  Thus one can evaluate the energy density of the primordial dark photons soon after the symmetry breaking as
\beq 
\rho_{\gamma'}^{\rm (pri)}\sim m_{\gamma'}\times n_{\gamma'}^{\rm (pri)} \sim e \frac{m_s}{\sqrt{\lambda}}\times n_{\gamma'}^{\rm (pri)}.
\eeq
By requiring this to be much smaller than $V_0$, we obtain
\begin{equation}
  e\gg 3\times 10^{-20} \sqrt{\beta}\lambda^{1/4}
\left(\frac{m_\phi}{10^{-15}{\rm \,eV}}\right)^\frac{1}{2}
\left(\frac{f_\phi}{10^{15}{\rm \, GeV}}\right)^{-\frac{1}{2}},  
\end{equation}
where we have used Eqs.\,\eqref{endnl} and \eqref{condition}, and $e |{\bm A}_{\rm end}|=m_\Psi$. In this case, the primordial dark photons become non-relativistic right after the trapping ends, and the abundance is negligibly small compared to the total DM density. 
In the region of our interest this is always satisfied.

Next we consider the dark photon production by the dark Higgs decay.
If the quartic coupling $\lambda$ is greater than $8 e^2$,
 the dark Higgs $s$ can promptly decay into dark photons.
 Due to the equivalence theorem, when the dark Higgs is much heavier than dark photons, the decay process can be described by the dark Higgs decaying into a pair of Nambu-Goldstone modes $\varphi = \sqrt{2} v \theta_\Psi$. The interaction is given by
\beq
\label{eqL}
\mathcal{L}\supset\frac{s}{\sqrt{2}v}(\del\varphi)^2, 
\eeq
and the decay rate is given by 
\beq
\Gamma_{s}(s\rightarrow\gamma'\gamma')\simeq\frac{m_s^3}{64\pi v^2}=\frac{\lambda m_s}{64\pi}\simeq 4.0{\rm meV}\cdot\lambda^{5/4},
\label{sdecay}
\eeq
where we have used (\ref{quartic}) in the third equality and assumed $m_s \gg m_{\gamma'}$. Unless $\lambda$ is taken to be extremely small, the decay rate is much larger than the Hubble parameter $H_{\rm end}$ at the end of the trapping, and so, the dark Higgs decays immediately.

In fact, the decay of the dark Higgs is further enhanced by the 
Bose factor. By analytically solving the Boltzmann equation with the Bose enhancement factor, we obtain the distribution function  for (longitudinal) dark photon at the redshift $z\lesssim z_c$ as~\cite{Moroi:2020has}, 
\beq
\label{fk}
f_k[z]= \frac{1}{2}\left(e^{32\pi^2 \frac{\Gamma_s n_s}{H m_s^3}}-1\right)\Theta\left(\frac{m_s}{2}-k\right)\Theta\left(k-\frac{m_s}{2} \frac{1+z}{1+z_{\rm c}} \right),
\eeq
where $k$ 
is the physical momentum of the dark photon, and $n_s$ is the number density of the dark Higgs. Here the exponent can be evaluated at $z_{c}$ by assuming that the decrease of $n_s$ due to the cosmic expansion is negligible compared to the production rate of the dark photon, i.e., we are focusing on a relatively short timescale so that $a \approx a_{\rm end}$. 
The exponent is inversely proportional to $H$ because the Bose enhancement is terminated when the produced dark photons are redshifted by the cosmic expansion. One can easily evaluate the exponent as $\sim \frac{\pi}{4} \lambda \frac{s^2_{\rm amp}}{m_s H}\sim \frac{\pi}{4} \frac{s^2_{\rm amp}}{v^2}\frac{m_s}{H} \sim 10^{25} \frac{s^2_{\rm amp}}{v^2}\frac{m_s}{0.01\,\rm eV}$, with $s_{\rm amp}$ being the effective oscillation amplitude of $s$, i.e., $s_{\rm amp}\equiv\sqrt{2n_s/(m_s)}$.\footnote{We note that this is the effective amplitude,  because  soon after the trapping ends the dark Higgs may experience tachyonic instability. However, this does not change our discussion because we use it to evaluate the order of magnitude of the exponent, and we solve the Boltzmann equation in terms of $n_s$, not the equation of motion for the $s$ condensate. In the case of the scalar condensate, it was shown that the exponent is smaller by a factor of 2 based on the QFT analysis~\cite{Moroi:2020bkq}.  See also the analysis in the narrow parametric resonance~\cite{Kofman:1997yn}.
} Soon after the trapping, this exponent is much larger than $1$ since $ s_{\rm amp}\sim v$. Thus the process would significantly reduces the number density of the non-relativistic $s$ by producing the ``laser" of the dark photon  until \cite{Nakayama:2021avl} $s_{\rm amp}\sim 10^{-13} v \sqrt{\frac{0.01{\,\rm eV}}{m_s}} $ when the exponent becomes of order unity.  
Then, the remnant of $s$ undergoes the usual perturbative decay, discussed above.

After the decay of dark Higgs, the produced dark photons are expected to be relativistic particles assuming $e^2/\lambda \ll 1/8$.
Then, the momentum of dark photons after the decay is given by
\beq
p_{\gamma'}(T)\simeq\frac{a_{\rm end}}{a}p_{\rm \gamma'}(T_{c}) \simeq \frac{a_{\rm end}}{2a}m_s,
\eeq
where the initial momentum of dark photons is approximately given by $m_s/2$.
We can estimate the timing when the dark photons become non-relativistic by $p_{\gamma'}(T_{\rm NR})\simeq m_{\gamma'}$, or equivalently,
\beq
a_{\rm NR}=\frac{m_s}{2m_{\gamma'}}a_{\rm end}\simeq100\left(\frac{m_s}{1\eV}\right)\left(\frac{m_{\gamma'}}{1\mu\eV}\right)^{-1},
\eeq
where the subscript `NR' means the variable is estimated when the dark photons become non-relativistic. For $e/\sqrt{\lambda} \lesssim 10^{-4}$, the dark photon remains relativistic till the present.

Assuming that the potential energy of the dark Higgs is instantaneously converted into dark photons, $\rho_{\gamma'}\simeq V_0=f_{\rm EDE}\rho_{\rm tot}(T_{c})$, the effective number of neutrinos of the dark photon is given by
\beq
\Delta N_{\rm eff}=\frac{\rho_{\gamma'}(T_{c})}{\frac{7\pi^2}{120}T_\nu^4(T_{c})}=\frac{4}{7}\left(\frac{11}{4}\right)^{4/3}f_{\rm EDE}g_{*0}\simeq0.37\left(\frac{f_{\rm EDE}}{0.05}\right),
\eeq
where $T_\nu$ represents the neutrino temperature. This satisfies the current constraint on the self-interacting dark radiation from the CMB observation, $\Delta N_{\rm eff}<0.6$~ \cite{Blinov:2020hmc}~(see the discussion below for the self-interacting properties.).

Let us  comment on the observational constraint on the dark photon abundance. For $e/\sqrt{\lambda} \gtrsim10^{-4}$, the produced dark photons become non-relativistic sometime until present, and afterwards they behave as self-interacting, hot DM. After the end of trapping, the dark photon do not free-stream due to the self-interaction (see below). Therefore, the damping effect of the small-scale structure is milder than free-streaming hot DM.
Also, the dark photon abundance is less than ${\cal O}(1)$\% of the observed DM density as far as $e^2/\lambda\lesssim1/8$ is satisfied, and the constraint from small-scale structure is considered to be irrelevant for our analysis.

Let us see that the dark photon radiation is self-interacting in the relevant era. The self-interaction of dark photon is obtained by integrating out the dark Higgs boson in \eqref{eqL}. We then find a term of 
\beq
{\cal L}_{\rm eff}\supset \frac{1}{v^2 m_s^2} (\partial^\mu \varphi \partial_\mu \varphi)^2 =  \frac{4}{V_0} (\partial^\mu \varphi \partial_\mu \varphi)^2.
\eeq 
This term equivalently represents the scattering of the longitudinal mode of the dark photon if the dark photon mass is small compared to the typical energy scale of interest. One can evaluate the typical scattering cross-section as  
\beq 
\sigma_{\gamma'\gamma'\to \gamma'\gamma'}\sim \frac{E_{\rm cm}^6}{4\pi^2 V_0^2}.
\eeq 
 Thus by assuming $\gamma'$ is around the equilibrium with a dark photon temperature $T_{\gamma'}$, the interaction rate, which is roughly the inverse of the free streaming length, is 
\beq 
\Gamma_{\rm th}\sim \frac{T_{\gamma'}^{9}}{4\pi V_0^2}. 
\eeq 
Since $T_{\gamma'} \sim (\frac{7}{12}0.37)^{1/4} T_\nu $, this rate is faster than the Hubble expansion rate at $z\gtrsim 1-10$. Thus, the produced dark photons are self-interacting dark radiation during and after the recombination era.

Lastly, we comment on our reference values $(f_{\rm EDE}, z_c)$ which were suggested in \cite{Poulin:2018cxd}.
In the original axion EDE model of Ref.~\cite{Poulin:2018cxd}, the axion changes from dark energy to dark radiation as it begins to oscillate. Since the equation of state oscillates violently with the axion oscillation, the CMB data and its prediction are compared under the fluid approximation by taking the time average of the equation of state. On the other hand, in our EDE model with dark Higgs, the equation of state actually changes from $-1$ to $1/3$ in a very short time because the dark Higgs decays to dark photons immediately after the end of the trapping. In this respect, the effects of the two EDE models on the CMB are expected to be very close, except for an oscillatory feature induced by the axion. Another slight difference is that in the axion EDE, the speed of sound is $1$ due to oscillations in the scalar field, while in the dark Higgs EDE, the speed of sound of the self-interacting dark photon is $1/\sqrt{3}$, which is slightly smaller.

\subsection{Allowed parameter region
\label{sec:region}}
Here let us summarize in \FIG{fig:mf} the above results. 
It shows the allowed parameter region for successful EDE as well as for explaining DM on the ($m_\phi, f_\phi$) plane. We have seven parameters, $v$, $\lambda$, $e$, $m_\phi$, $f_\phi$, $\beta$, and $\theta_*$. This figure is drawn based on the assumption that the potential energy of the dark Higgs corresponds to a small fraction of the total energy, $f_{\rm EDE}=0.05$ at $z_{c}=5000$, and $H_{\rm osc}=m_\phi$. Among the model parameters, we fix $\beta$, $\theta_*$, and $\lambda$. 
In particular, we take $\lambda=1$, $\theta_*=1$, and $\beta=100$ in \FIG{fig:mf}.
The black solid line denotes $\Omega_\phi=\Omega_{\rm DM}$, above which the axion is overproduced than DM, represented by the light red shaded region. On the red dot-dashed line the axion constitutes $1\%$ of DM. The gray shaded region is excluded because the decay of dark Higgs is kinematically forbidden. The gray dashed contour lines represent the gauge coupling constant. 
In the region above the blue dotted line, the axion has the longer lifetime than the cosmic age. The small cyan region represents where it is constrained by the bound on decaying DM by (\ref{lifetime}).
(See Ref.~\cite{Alvi:2022aam} for the bound on the case when a fraction of DM decays.) One can see that the dark Higgs EDE works well in a very wide parameter range.

\begin{figure}[t!]
\includegraphics[width=11cm]{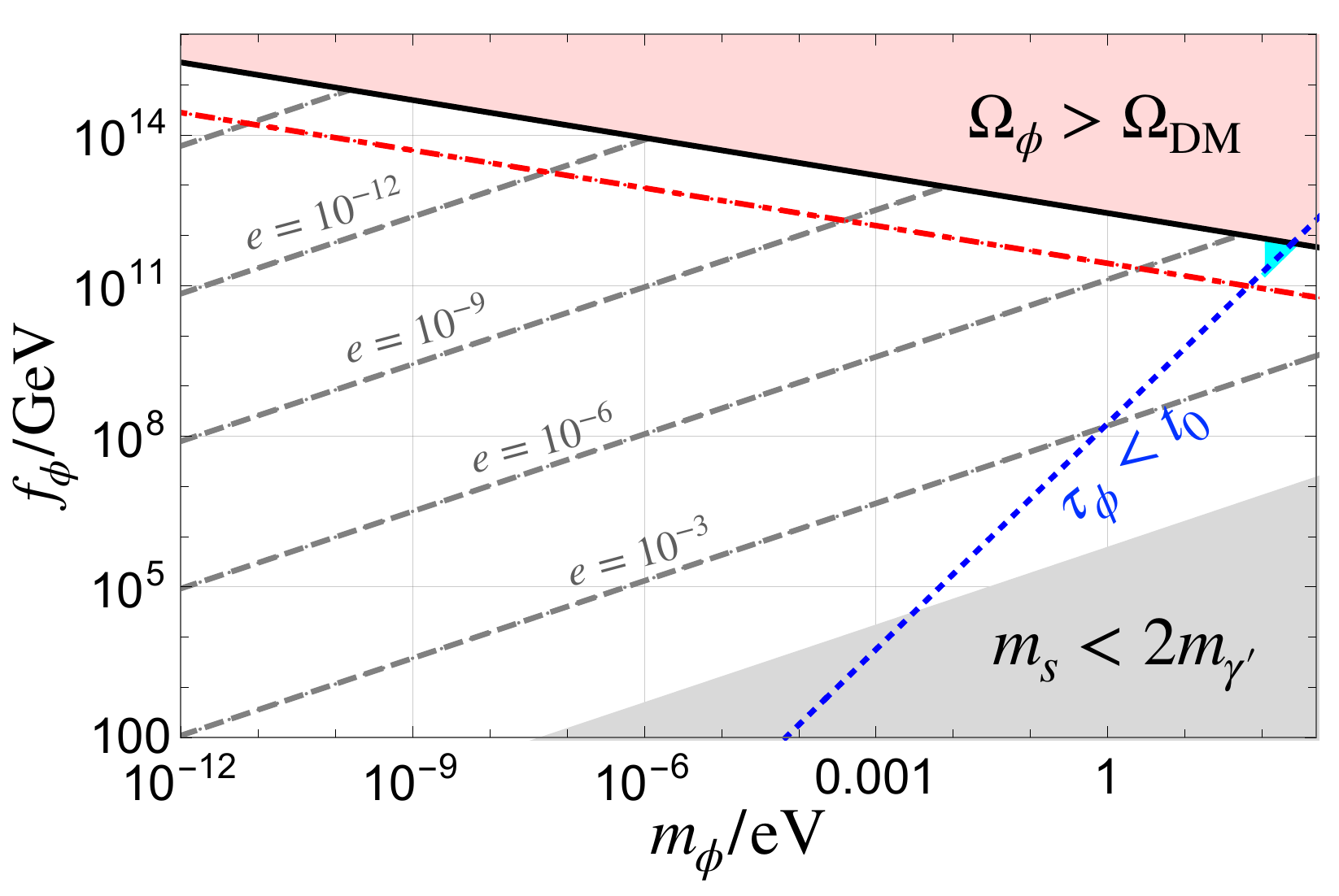}
\centering
\caption{The parameter region allowed for our scenario on the plot $(m_\phi,f_\phi)$ with $\lambda=1$. We take $\theta_*=1$ and $\beta=100$. The black solid line denotes $\Omega_\phi=\Omega_{\rm DM}$, above which the axion is overproduced represented by the light red shaded region, and the red dot-dashed line denotes $\Omega_\phi=0.01\Omega_{\rm DM}$. The gray shaded region represents the regions excluded by no Higgs decay ($m_s>2m_{\gamma'}$). In the region above the blue dotted line, the axion has the longer lifetime than the cosmic age, and the constraint from the decaying axion into dark radiation can be applied, denoted by the cyan shaded region. The gray dashed contour lines denote the gauge coupling constant.}
\label{fig:mf}
\end{figure}

The result for $\lambda=10^{-6}$ is shown in \FIG{fig:fae2}, and compared to Fig. 1, the limit on the lifetime of the axion DM remains the same, but $e$ is smaller.
This is because the potential of the dark Higgs becomes flatter so that it would be trapped for a longer time for the same $e$. This results in the smaller value of $e$ for the trapping to 
end around $z_c = 5000$.

Figs.~\ref{fig:mf} and \ref{fig:fae2} also show that axion can explain all DMs for a wide range of axion masses on the black solid line. Interestingly, part of this mass range overlaps with the mass range explored by various axion DM search experiments. If the axion is coupled to SM particles (especially photons), then axions could be searched for by these experiments. This point will be discussed in \SEC{sec:conclusion}. Note that the lighter the axion mass, the smaller the gauge coupling $e$ and the larger the hierarchy between $\beta$ and $e$, which can be explained by the UV model described in the \SEC{sec:model}.

\begin{figure}[t!]
\includegraphics[width=11cm]{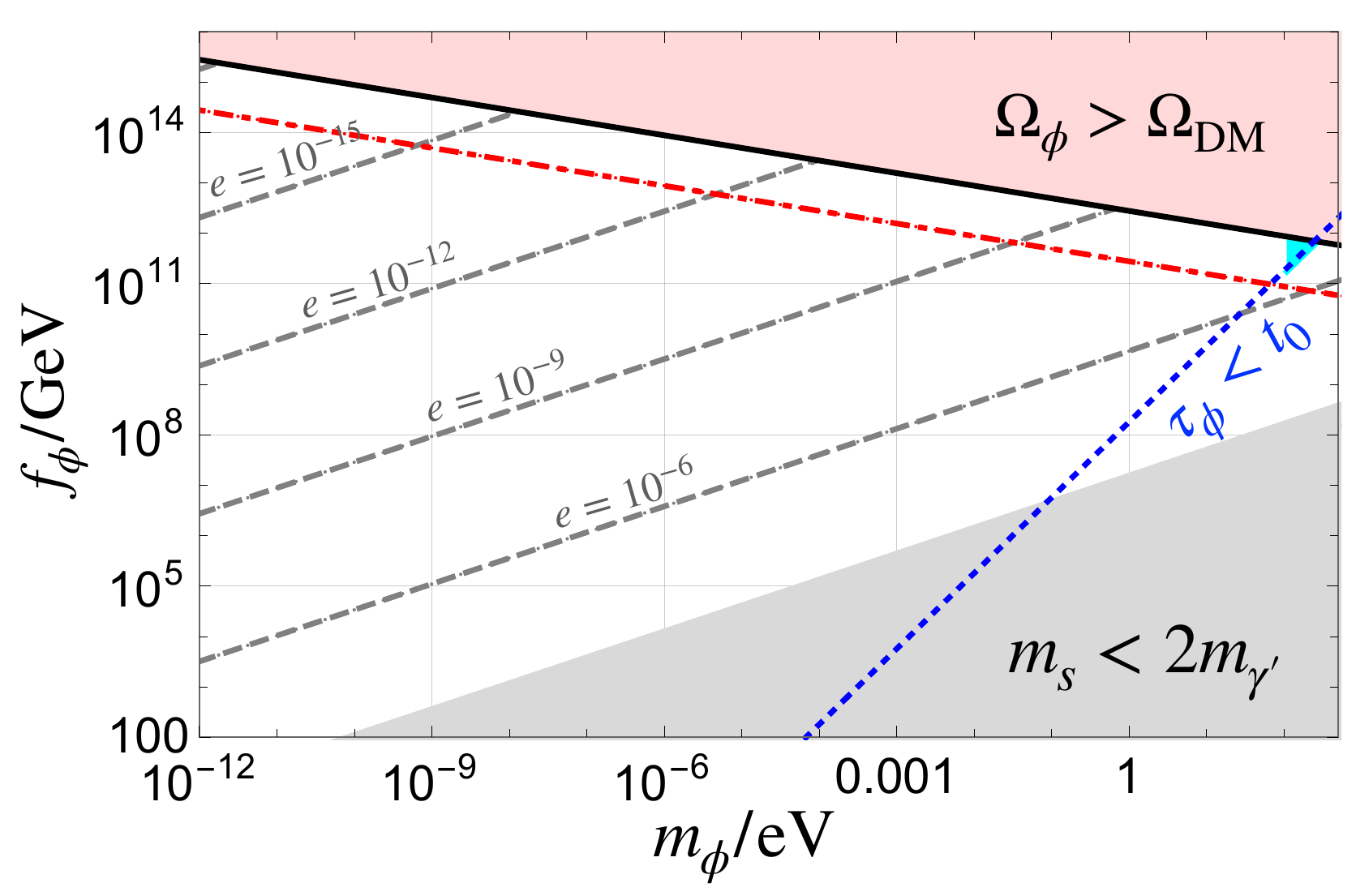}
\centering
\caption{Same as Fig.~\ref{fig:mf} except for $\lambda=10^{-6}$. 
We take $\theta_*=1$ and $\beta=100$. 
}
\label{fig:fae2}
\end{figure}

\section{Possible extensions
\label{sec:model}}

\subsection{Two hidden U(1) gauge symmetries with kinetic mixing} \label{sec:mixing}
In \SEC{sec:region} we have found that the dark Higgs EDE works well, but we need a relatively small gauge coupling $e$ while the axion-dark photon coupling is rather large as $\beta\sim\mathcal{O}(10-100)$. In particular, the axion DM requires $e\lesssim10^{-7}$ for $\beta = 100$ (see Fig.~\ref{fig:mf}).
This is because the axion requires a relatively large decay constant to explain all DM, which results in the very strong trapping of dark Higgs at the origin. Since the end of trapping is determined by $e|\bm{A}_{\rm end}|\simeq m_\Psi$, we need to take a special value of $e$ so that the trapping regime terminates just at the right timing, $z_c \approx 5000$.

Here we show below two ways to explain this hierarchy. The first one is to consider the aligned or  clockwork axion model \cite{Kim:2004rp,Choi:2014rja,Higaki:2014qua,Higaki:2015jag,Choi:2015fiu,Kaplan:2015fuy,Giudice:2016yja,Higaki:2016yqk,Farina:2016tgd}. 
It was first pointed out in Ref.~\cite{Higaki:2016yqk} that the axion coupling to dark gauge bosons can be naturally enhanced by many orders of magnitude in the clockwork mechanism. In this set-up, there are $N$ axions, and $N-1$ of the corresponding global U(1) symmetries are explicitly broken by a certain special combinatorial pattern, which give masses to $N-1$ axions and leave one combination massless.
This massless direction is  identified with the axion in our set-up, and its effective decay constant is exponentially larger than the characteristic symmetry breaking scale of the U(1) symmetries. One can give a small mass to the axion by including another explicit breaking. Since the coupling to dark photons does not change the structure of the mass matrix for the axions, the lightest axion can have a rather strong coupling to dark photons (or a natural size in terms of the original symmetry breaking scale), when compared to its enhanced decay constant~\cite{Higaki:2015jag}.   The drawback of this scenario might be that the UV completion is somewhat complicated.

As another simple explanation of the hierarchy between $e$ and $\beta$, let us consider a  model with  two hidden U$(1)_{\rm H}$ symmetries, U$(1)_{\rm H1}$ and U$(1)_{\rm H2}$. We assume that the dark Higgs is charged under U$(1)_{\rm H1}$, while the axion is coupled to U$(1)_{\rm H2}$:
\beq
\mathcal{L} &=& (D_\mu\Psi)^\dag D^\mu\Psi-V_\Psi(\Psi,\Psi^\dag)-\frac{1}{4}X_{1\mu\nu}X_1^{\mu\nu}-\frac{1}{4}X_{2\mu\nu}X_2^{\mu\nu}-\frac{\chi}{2}X_{1\mu\nu}X_2^{\mu\nu}\nonumber\\
&+&\frac{1}{2}\del_\mu\phi\del^\mu\phi-V_\phi(\phi)-\frac{\beta}{4f_\phi}\phi X_{2\mu\nu}\tilde{X}_2^{\mu\nu},
\label{Lagrangian2}
\eeq
where $X_{i}^{\mu\nu}\equiv\del^{\mu}A_{i}^{\nu}-\del^{\nu}A_{i}^{\mu}$ denotes the field strength tensor for the U$(1)_{{\rm H}i}$ with $i=1,2$, $D_\mu=\del_\mu-ie_1A_{1\mu}$ with $e_1$ the gauge coupling constant of the U$(1)_{\rm H1}$, $\chi$ is the kinetic mixing, and $\beta$ is defined as the axion coupling to the dark photon of U(1)$_{\rm H2}$. The important point is that $e_1$ determines the dark photon mass coupled to the dark Higgs, and it needs to be taken small (but larger than the original set-up, see below), while $e_2$ can be of order unity.\footnote{One still needs to enhance the axion-dark photon coupling by a factor of ${\cal O}(10-100)$, which can be easily explained by the clockwork mechanism with a few extra axions.} 
The natural size of the the kinetic mixing is $\chi \sim  0.01 e_1e_2$, and this is indeed the case if there are bi-charged particles in the UV theory.
Through the small kinetic mixing $\chi$, 
the dark Higgs is also coupled to  the dark photon of U(1)$_{\rm H2}$, which is produced from the axion by the tachyonic instability.
 To see this explicitly, let us perform the following linear transformation,
\beq
A'_{1\mu}&=&A_{1\mu}+\chi A_{2\mu}\nonumber\\
A'_{2\mu}&=&\sqrt{1-\chi^2}A_{2\mu}.
\eeq
The covariant derivative now reads $D_\mu\simeq\del_\mu-ie_1A'_{1\mu}+ie_1\chi A'_{2\mu}$. In this case, the dark Higgs has a minicharge $e_1\chi$ under U$(1)_{\rm H2}$ and can be trapped by the effective mass $e_1\chi|\bm{A}'_2|$ with a large $\beta$ and a tiny effective gauge coupling. Note that the gauge coupling $e_1$ does not have to be as small as in the original set-up, because the effective mass of the dark Higgs is also suppressed by the kinetic mixing. 
When the dark Higgs develops a nonzero VEV at the end of the trapping, the U$(1)_{\rm H1}$ gets spontaneously broken, but there remains a massless dark photon which is  the dark photon of the U$(1)_{\rm H2}$. Performing a linear transformation so that the mass eigenstate is $A_{1\mu}$, one can see that the massive dark photon for the U$(1)_{\rm H1}$ also couples to the axion in vacuum.

For the dark Higgs EDE to work well in this two U$(1)_{\rm H}$ model, the dark Higgs must decay into dark radiation. Since the dark Higgs couples only to the massive dark photon after the spontaneous breaking of U$(1)_{\rm H1}$, it promptly decays as in (\ref{sdecay}). 
Although the massive dark photon decays into the three massless dark photons via the kinetic mixing according to the Furry's theorem, its lifetime is extremely long, and the decay does not change our discussion in the original set-up.

We show the allowed region in \FIG{fig:double}. We take $\lambda=1$, $\theta_*=1$, $e_2=1$, and $\beta=100$, assuming $\chi=e_1e_2/16\pi$. One can see that the required size of $e_1$ becomes larger compared to Fig.~\ref{fig:mf}.
The axion DM can be explained for $e_1\lesssim10^{-2}$, and there is only mild hierarchy between $e_2$ and $\beta$.

\begin{figure}[t!]
\includegraphics[width=11cm]{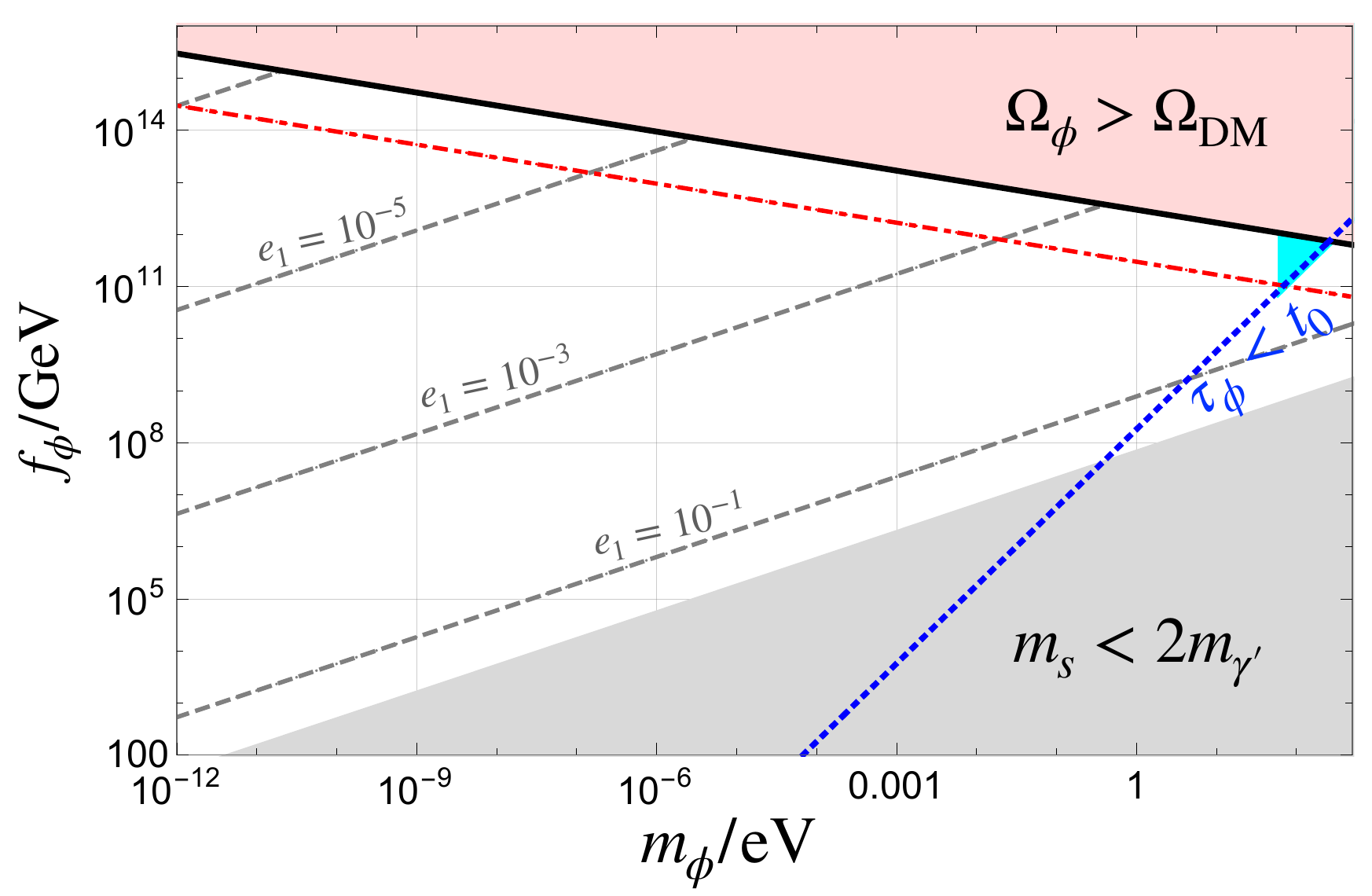}
\centering
\caption{The viable parameter region in the two  U$(1)_{\rm H}$ model. We set $\lambda=1$, $\theta_*=1$, $e_2=1$, and $\beta=100$, assuming $\chi=e_1e_2/16\pi$. The reader should refer to the caption of \FIG{fig:mf} for the explanation of the shaded regions and lines.
}
\label{fig:double}
\end{figure}

\subsection{QCD axion}

Let us discuss the possibility that the QCD axion, $a$, plays the role of the axion in the previous section. 
The QCD axion is the leading candidate that solves the strong CP problem~\cite{Peccei:1977hh,Peccei:1977ur,Weinberg:1977ma,Wilczek:1977pj}. In addition it is one of the most plausible candidates for  DM. The feasible range of the decay constant $f_a$ to account for DM by the axion generated by the misalignment mechanism~\cite{Preskill:1982cy,Abbott:1982af,Dine:1982ah} is wider than naively thought, and it is given by $f_a= 10^{9}-10^{18}$ GeV. This is because the initial misalignment angle depends on the inflation scale and inflaton couplings. In particular, the small initial angle required for large $f_a$ can be achieved naturally by considering low inflationary scales~\cite{Graham:2018jyp, Guth:2018hsa}, and the initial angle near $\pi$ required for small $f_a$ can be achieved by 
shifting the axion to the potential maximum~\cite{Takahashi:2019qmh} (the idea to put an axion-like particle on the hilltop was proposed in \cite{Daido:2017wwb}, see also \cite{Co:2018mho, Huang:2020etx} with other fields introduced to flip the sign of the axion potential).
If one can identify the axion with the QCD axion in our scenario, the QCD axion abundance is
reduced by $\mathcal{O}(10)$ for $\beta=O(10-100)$ via tachyonic production of dark photons compared with the conventional estimation~\cite{Kitajima:2017peg} (see also \cite{Agrawal:2017cmd}). Thus the allowed range of the decay constant can be slightly enlarged.

The main difference from the discussion so far is that the QCD axion has a temperature dependent mass:
\begin{align}
\label{mass}
m_a(T) \simeq
\begin{cases}
\displaystyle{\frac{\sqrt{\chi_0}}{ f_a}} \left(\frac{T_{\rm QCD}}{T}\right)^n &~~T \gtrsim T_{\rm QCD}\\
\displaystyle{5.7 \times 10^{-6} \left(\frac{10^{12}{\rm GeV}}{f_a}\right)  {\rm eV}}&~~ T\lesssim  T_{\rm QCD}
\end{cases},
\end{align}
with $n \simeq 4.08$~\cite{Borsanyi:2016ksw},  $T_{\rm QCD}\simeq 153$meV and $\chi_0 \simeq \left(75.6 {\rm MeV}\right)^4$. 
The onset of oscillation of the QCD axion is at $T=T_{\rm osc}\sim 1\,$GeV, which is lower than the axion with the same mass as the QCD axion at zero temperature.

Compared to the axion with a temperature-independent mass, the dark photon produced from the QCD axion is considered to have the following two differences. First, compared to the axion with the same (constant) mass, the QCD axion starts to oscillate at a later time. Therefore, the amplitude of dark photons is larger. On the other hand, the mass of the QCD axion increases between the onset of oscillation and the nonlinear regime of the system, so that the axion oscillation amplitude becomes smaller than its initial value. Therefore, the amplitude of the produced dark photon also becomes smaller. While one needs lattice calculations to precisely evaluate these effects,  the first of the two effects is expected to be stronger than the second because of the larger abundance of QCD axion compared to the axion of the same mass. That is, the effect of trapping dark Higgs is expected to be stronger in the case of QCD axion. Therefore, a smaller gauge coupling constant is needed to terminate the trapping at just the right time.

We comment on that a larger $e$ but smaller $\beta$ may also be consistent with the scenario. In particular, to explain the axion DM with $f_\phi \sim 10^{9-10}\,$GeV, the initial misalignment angle should be close to $\pi$. In this case, we expect further enhancement of the dark photon production due to the anharmonic effect, and a smaller $\beta$ may lead to the dark photon amplitude for the trapping.
Also, considering the rather strong trapping effect, one may consider $\beta = {\cal O}(1)$ or smaller for which the system does not enter the nonlinear regime, but the produced dark photons may be able to trap the dark Higgs for a larger gauge coupling $e$.
This argument is not limited to the case of the QCD axion, but also applies to the axion with the constant mass.

{Alternatively, one may consider the heavy QCD axion whose mass is heavier than \eqref{mass} due to the small instantons. Then the temperature-dependence of the axion mass is significantly reduced, and the gauge coupling constant can be larger.
For instance, we can extend the color sector to $SU(3)\times SU(3)\times SU(3)$ with two additional axions below the Peccei-Quinn breaking scale as studied in Refs.\,\cite{Agrawal:2017ksf,Csaki:2019vte,Takahashi:2021tff}. Two bifundamental Higgs fields spontaneously break $SU(3)^3\to SU(3)$. We obtain a sizable small instanton contribution to the lightest axion, which solves the strong CP problem. The resulting axion potential has the CP conserving minimum, and the QCD axion mass is heavier. Since the heavy QCD axion mass is (almost) independent of the temperature, the resulting parameter region is essentially the same as the one shown in the previous section, {except for that we are limited to the region where the axion mass is heavier than the usual QCD axion mass (\ref{mass}).}}

\subsection{(Un)naturalness of the dark Higgs mass}
Let us now discuss the potential fine-tuning of the dark Higgs sector and its relaxation, noting that the mass of dark Higgs is very small. This can be explicitly seen when we introduce a certain ``Peccei-Quinn" fermion $f$ of ${\rm U}(1)_{\rm H2}$ charge $q_2$ in \SEC{sec:mixing} to induce the axion-dark  photon coupling. It is expected that,  by integrating out the fermion, there arises a radiative correction to the dark Higgs mass of order \beq \delta m_\Psi^2 \sim \frac{e_1^2 \chi^2 e_2^2}{(16\pi^2)^2}  q_2^2 m_f^2\eeq
with $m_f$ being the mass of the fermion.
By requiring the mass correction to be smaller than $1$\,eV, $m_f$ should be comparable to or smaller than TeV for the parameter region in Fig. \ref{fig:double}. 
If this condition is satisfied,  the model may become technically natural. However, to satisfy the 't Hooft naturalness, one has to relate the small dark Higgs mass with symmetry. This could be realized in a supersymmetric UV completion.

\section{Discussion and conclusions
\label{sec:conclusion}}
Let us discuss the experimental and observational implications of the axion coupling to the SM  photons. Such axions coupled to photons are often called axion-like particles (ALPs). The interaction is given by
\beq
\mathcal{L}_{a\gamma}= 
-\frac{g_{\phi\gamma}}{4}\phi F_{\mu\nu}\tilde{F}^{\mu\nu}
=-\frac{\beta_{\rm EM}}{4f_\phi}\phi F_{\mu\nu}\tilde{F}^{\mu\nu},
\eeq
where $F_{\mu\nu}$ is the field strength tensor for photons, $\tilde{F}_{\mu\nu}$ is its dual, and $\beta_{\rm EM}$ and $g_{\phi\gamma}$ denote the axion coupling with photons. We show the allowed region with the current bounds on and the future sensitivities to the axion-photon coupling in \FIG{fig:coupling}.  
We take $\lambda=1$, $\beta=100$, $\beta_{\rm EM}=1/100$, and $\theta_*=1$. The colored shaded regions denote the current bounds on $|g_{\phi\gamma}|$, and the dashed lines denote the future sensitivities. We refer to the summary for the axion-photon coupling limits \cite{AxionLimits} and references therein. 
We note that some constraints and sensitivities are based on the assumption that the axion saturates the observed DM abundance.
The axion with $m_\phi\lesssim10^{-5}\eV$ that can explain all DM will be probed by various experiments, such as ADMX \cite{Stern:2016bbw}, FLASH \cite{FLASH}, DM-Radio \cite{DMRadio:2022pkf}, and SRF \cite{Berlin:2020vrk}.
The parameter region where the axion is subdominant DM can  also be probed if the sensitivity reaches are below the black solid line.
However, the axion abundance would be altered by how the axion DM is created in the early universe. For instance, while we take $\theta_*=1$ here, the axion abundance can be enhanced by the anharmonic effect, but we cannot take too small an initial angle because we need $\beta\theta_*\sim\mathcal{O}(10-100)$ for the efficient tachyonic production of the dark photon. In this case the contour of $e$ may be slightly altered.

With $\beta \theta_*\lesssim 1$, we can still produce the abundant primordial dark photon with a relatively light axion (see Ref.\,\cite{Alonso-Alvarez:2019ssa} for the relation between the particle picture and the tachyonic instability in the axion-photon system). In this case, 
we can  study the primordial dark photon production from the axion bose-enhanced decay, following an equation similar to Eq.\,\eqref{fk} with  $\Gamma_{s\to \gamma'\gamma'}$ and $m_s$ replaced by $\Gamma_{\phi \to \gamma'\gamma'}$ and $m_\phi$, respectively.
Then, we expect that the non-thermal trapping is possible by the produced dark photon with  larger $e$.

\begin{figure}[t!]
\includegraphics[width=13cm]{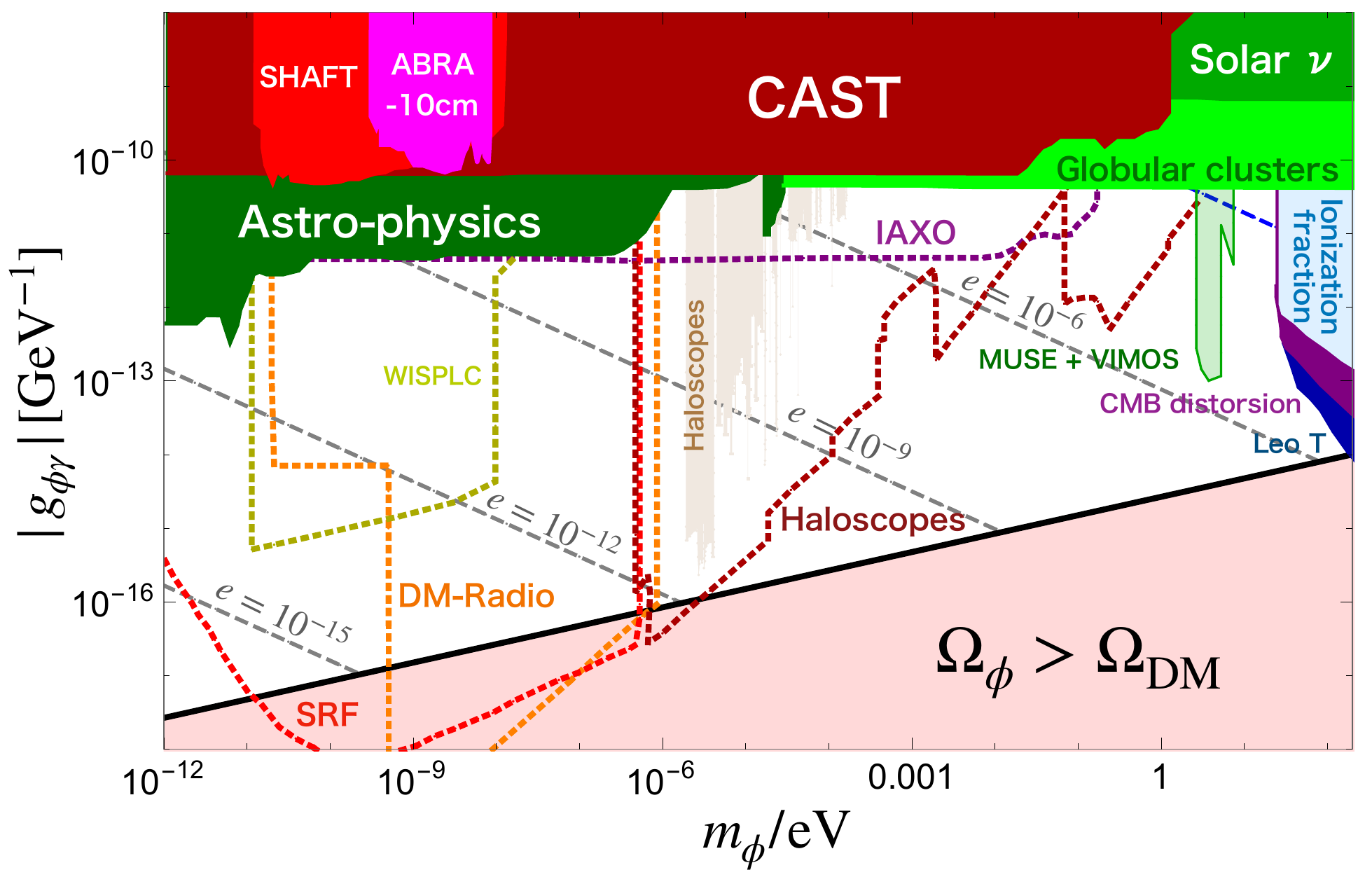}
\centering
\caption{
Current limits and future experimental sensitivity to the ALP coupling to photons, and the ALP DM abundance and predicted dark U(1) gauge coupling constant in the dark Higgs EDE scenario.
 We take $\lambda=1$, $\beta=100$, $\beta_{\rm EM}=1/100$, and $\theta_*=1$.} 
\label{fig:coupling}
\end{figure}

So far, we have focused on the light axion that is stable on cosmological times scales. Here we comment on the possibility of a heavy, unstable axion which decays after the tachyonic production of dark photons. In this case, the axion decays into  dark photons at a certain time that depends on $m_\phi$ and $f_\phi$. The cosmological impacts of such unstable axion depend strongly on the axion abundance at the time of the decay. As long as the axion decays into dark photons, there is almost no noticeable effect if its abundance is less than a few percent of the total energy of the universe at the time of decay.  If the axion accounts for more than several percent of DM and completely decays after recombination until present, this would be inconsistent with the CMB observations~\cite{Enqvist:2015ara,Enqvist:2019tsa,Alvi:2022aam}. 
Also, the amount of dark photons after the decay should not exceed the limits from CMB observations on the abundance of dark radiation~\cite{Planck:2018vyg}. While such unstable axion cannot explain DM, it may be possible to induce the dark Higgs EDE if it decays in the early universe and if its abundance is small enough. One motivation for considering such a heavy axion is that it could be a candidate for the inflaton if only a plausible reheating process could be devised. 
Note that the EDE favors the spectral index of the primordial power spectrum larger than in the $\Lambda$CDM case.
The cosmological implications of the larger $n_s$ suggested  in some of the solutions to the Hubble tension were discussed in Refs.~\cite{Takahashi:2021bti,DAmico:2021fhz,Kallosh:2022ggf,Lin:2022gbl,Ye:2022efx,DAmico:2022agc}.
Since our scenario does not rely on the details of the inflation model, we will not go there in this paper.

So far, we have focused on the parameter region $e^2 \ll \lambda/8 $ so that the dark Higgs boson after the trapping decays into the dark photons. Now, let us consider the case
of 
\beq 
\label{ineq} 
e^2\gg {\lambda}.
\eeq 
While this regime may not be viable in the context of EDE,\footnote{If the first-order phase transition happens at around the recombination, in the context of the early dark energy,  the CMB anisotropy constraint would be severe unless the bubble size is extremely small (see c.f. Ref.~\cite{Niedermann:2019olb}).} it is interesting to study because it may provide an observable prediction of the non-thermal trapping scenarios. 
We show below that, unlike the discussion so far, a first order phase transition generically happens in this regime.
As we have mentioned, by slowly varying $s$ we find that the primordial dark photon amplitudes change via $|{\bm A}|\propto 1/({k^2+ m^2_{\gamma'}})^{1/4}$, where $k$ is the characteristic physical momentum of the dark photon.
This is because $|{\bm A}|^2\sqrt{k^2+ m^2_{\gamma'}} $ is an adiabatic invariant in the flat spacetime limit, when the system changes with a time scale longer than 
$
\Delta t_{\rm adiabatic}=\frac{2\pi}{\sqrt{k^2+m^2_{\gamma'}}}.
$
This implies that the effective potential for the dark Higgs potential has the following form,
\beq 
V_{\rm eff}= V_\Psi +  \frac{|{\bm A}_0|^2 k}{\sqrt{k^2 +e^2 |\Psi|^2}} e^2 |\Psi|^2,
\eeq 
where ${\bm A}_0$ represents the spatially averaged field value of the dark photon at $|\Psi|=0$.
We can see that the second term is approximately a linear potential at $|\Psi| > k/e$.  
In this case, if 
\beq 
|{\bm A}_0|^2 k e v  \ll m_\Psi^2 v^2
\eeq 
around the end of the trapping, there is a potential barrier between the origin and the true minimum.
By taking $|{\bm A}_0|^2e^2=|{\bm A}_{\rm end}|^2e^2=m_\Psi^2$, 
this condition is reduced to
\beq k\ll m_{\gamma'}.\eeq 
Namely, if the primordial dark photons are non-relativistic after $s$ settles to its true minimum, the linear term plays an important role to make the potential barrier.
The end of the trapping is then triggered by quantum tunneling, i.e., a first-order phase transition happens.
 
The reason why we imposed \eqref{ineq} is because the time scale of the tunneling $\Delta t_{\rm Higgs}\sim 1/m_s$ should be much longer than $\Delta t_{\rm adiabatic}$, i.e., $\Delta t_{\rm Higgs}\gg \Delta t_{\rm adiabatic}\sim 1/m_{\gamma'}$ which leads to \eqref{ineq}. 
Thus the non-thermal trapping scenario with \eqref{ineq} provides a new possibility of the strong first order phase transition and the subsequent production of gravitational waves. Indeed the potential with the effective linear term may be  generic if the number of light particles trapping the dark Higgs field is conserved during the phase transition. For instance, it is possible that weakly coupled light particles cannot annihilate in the time scale that $s$ evolves.

The first order phase transition in the scalar trapped by the number-conserving particles could be compared with the phase transition of the thermal inflation~\cite{Yamamoto:1985rd,Lyth:1995ka} where the flaton potential also has a barrier. The phase transition was shown to proceed via a phase-mixing but not via a tunneling~\cite{Hiramatsu:2014uta} (see also \cite{Easther:2008sx}).  This is because the thermal fluctuation is so large that the flaton easily jumps over the potential barrier before the tunneling to occur. In our case of the non-thermal trapping, on the other hand, there is no sizable thermal fluctuation, and, in addition, the potential barrier does not need to be very close to the origin for the tunneling to occur. Thus, the first-order phase transition is considered to be generic in such  non-thermal trapping or a trapping by number-conserving particles. \\

In this paper we have proposed a novel scenario in which the dark Higgs trapped at the origin explain the EDE, one of the plausible solutions to the Hubble tension. The axion condensate triggers tachyonic production of dark photons, which keep the dark Higgs trapped for a long time. We have identified the viable model parameters where the Hubble tension is solved by EDE and the axion explain all DM for the axion decay constant in the intermediate scales 
 and a wide range of the axion mass.
We have also shown that it is possible identify the axion with the QCD axion.

\section*{Acknowledgments}
We thank Naoya Kitajima for useful communications.
The present work is supported by the Graduate Program on Physics for the Universe of Tohoku University (S.N.), JST SPRING, Grant Number JPMJSP2114 (S.N.), JSPS KAKENHI Grant Numbers  20H01894 (F.T.) 20H05851 (F.T. and W.Y.), 21K20364 (W.Y.), 22K14029 (W.Y.), and 22H01215 (W.Y.)
and 
JSPS Core-to-Core Program (grant number: JPJSCCA20200002) (S.N. and F.T.).

\clearpage
\appendix
\label{app}
\section{Estimates for the start of nonlinear regime
\label{sec:app}}
This section is devoted to the explanation for how much dark photons are produced by the tachyonic instabilities till the beginning of the nonlinear regime. To this end, we study when the energy density of dark photons becomes comparable to that of the axion and the system enters the non-linear regime. One can see from the equation of motion (\ref{eq:Apm}) that the dominant growing mode is $k_{\rm{peak}}/a \sim \beta |\dot\phi|/(2f_\phi) \sim \beta m_\phi |\phi|/(2f_\phi)$, where $|\phi|$ denotes the oscillation amplitude. It takes the maximal value, $k_{\rm{peak}}/a \sim \beta m_\phi\theta_*/2$, at the onset of the axion oscillations with $\theta_*$ the initial amplitude, and it gradually decreases proportional to the oscillation amplitude. 
The efficient tachyonic production of dark photons requires $\beta \theta_* = {\cal O}(10-100)$. The non-linear regime begins soon after the system satisfies
\beq
\frac{1}{2}m_\phi^2f_\phi^2\theta_*^2\left(\frac{a_{\rm{osc}}}{a_{\rm nl}}\right)^3\simeq \frac{k_{\rm{peak}}^2(t_{\rm{nl}})}{2a_{\rm nl}^2}|{\bm{A}_{\rm nl}}|^2,
\label{nl}
\eeq
where we have approximated the energy of dark photons to the gradient energy of the dominant growing mode. Thus, the field value of the dark photon can be estimated as
\beq
|\bm{A}_{\rm nl}| \simeq \frac{2f_\phi}{\beta}.
\label{Anl}
\eeq

Instead of considering the equilibrium between the energy densities of the dark photon and the axion, there is another way to estimate when the backreaction on the axion dynamics becomes significant. In the equation of motion for the axion (\ref{axioneom}), there are two terms coming from the axion potential and the coupling to the dark photons. The backreaction on the axion dynamics is considered to be significant when the latter dominates over the former. In \cite{Kitajima:2021bjq}, the authors numerically confirmed that, since the axion oscillation amplitude decreases with time, the timing evaluated by this method is consistent with that evaluated by comparing the energy densities of axion and dark photons.

Note that we have assumed here that the backreaction of the dark photon production is not significant during a single oscillation of the axion, before the non-linear regime. This is because we adopt a mildly enhanced axion coupling, $\beta \theta_* = 100$. On the other hand, if it were larger than ${\cal O}(10^2)$ or so, one cannot the neglect the backreaction on the axion dynamics even during a single oscillation, and one should treat the effect as a frictional force on the axion motion~\cite{Kitajima:2017peg}. In fact, it was shown in Ref.~\cite{Kitajima:2017peg} that the QCD axion abundance can be enhanced for  such a large coupling due to the extra frictional force. In the present case, for the parameters we adopted, it is known \cite{Kitajima:2021bjq} that the axion oscillates many times before the non-linear regime begins, and such a linear analysis is justified until the nonlinear regime sets in.

We also estimate the exponential enhancement factor from the onset of the axion oscillation until the beginning of the non-linear regime. We obtain the exponential growth factor of the mode with the wave number $k_{\rm{peak}}/a$ as
\beq \label{eq:growth}
\exp\left(\frac{1}{2}\int^{t_{\rm nl}}_{t_{\rm osc}} dt\frac{\beta m_\phi|\phi(t)|}{2f_\phi}\right) &\simeq& \exp\left[\frac{\beta m_\phi\phi_{*}}{2\pi f_\phi} \int^{t_{\rm nl}}_{t_{\rm osc}} dt \left(\frac{a(t)}{a_{\rm{osc}}}\right)^{-\frac{3}{2}} \right],\nonumber\\
&\simeq& \exp\left[\frac{\beta\theta_*}{\pi}\left(\left(\frac{a_{\rm nl}}{a_{\rm osc}}\right)^{1/2}-1\right)\right],
\eeq
where the oscillatory part of $\phi(t)$ is replaced with the averaged value, and we used the relation, $H_{\rm osc} = 1/(2 t_{\rm osc}) \simeq m_\phi$.  Note that $\frac{1}{2}$ in the first term means that the enhancement of each helicity mode is switched every half a period. The initial field value of dark photon is roughly given by $|\bm{A}_{\rm{osc}}| \sim k_{\rm{peak}}/a_{\rm{osc}}$, and thus, from Eqs.~(\ref{Anl}) and (\ref{eq:growth}), we obtain
\beq
\frac{a_{\rm nl}}{a_{\rm osc}} \sim \left[1+\frac{\pi}{\beta\theta_*}\ln\left(\frac{4f_\phi}{\beta^2 \theta_* m_\phi}\right)\right]^2.
\label{nlosc}
\eeq
Note that since the dominant growing mode changes with time due to the cosmic expansion, one needs numerical simulation to calculate precisely when the non-linear regime begins. The above estimate is in a good agreement with the numerical result within a factor of $O(1)$, when the growth rate of the instabilities is sufficiently fast \cite{Kitajima:2021bjq}. On the other hand, if the growth rate is relatively small due to large hierarchy between $f_\phi$ and $m_\phi$ and/or small $\beta$, the above estimate breaks down because the instability band becomes narrow.

\section{Additional mass of dark Higgs from the Higgs portal coupling}
\label{app:trapping}
As a possible way to keep the dark Higgs at the origin for the axion mass lighter than $0.1$\,eV, we introduce a Higgs portal coupling \cite{Patt:2006fw},
\beq
\mathcal{L}_{\rm portal}=-\lambda_{H\Psi}H_s^\dagger H_s\Psi^\dag\Psi,
\eeq
where $\lambda_{H\Psi}$ is a portal coupling constant and $H_s$ is the SM Higgs doublet. This model is applicable to both the matter dominated and radiation dominated universe, but we assume the latter one for simplicity. The portal coupling constant is roughly estimated as
\beq
\lambda_{H\Psi}\sim\left(\frac{m_\Psi}{v_{\rm EW}}\right)^2\simeq10^{-23}\left(\frac{m_\Psi}{1\eV}\right)^2,
\eeq
where $v_{\rm EW}\simeq256\GeV$ is the VEV of the SM Higgs. This is because the SM Higgs VEV contributes to the dark Higgs mass as $\sqrt{\lambda_{H\Psi}v_{\rm EW}^2}$ and we adopted the largest value that does not require cancellation between the contributions to the dark Higgs mass. Such a tiny coupling does not allow the dark sector to be thermalized, but the dark Higgs acquires a thermal effective mass through the portal coupling,
\beq
\Delta m(T_{\rm osc}> 100\,{\rm GeV})\sim\sqrt{\lambda_{H\Psi}T^2_{\rm osc}}\sim100\eV\left(\frac{m_\Psi}{1\eV}\right)\left(\frac{m_\phi}{1\eV}\right)^{\frac{1}{2}},
\eeq
where $T_{\rm osc}\simeq\sqrt{ m_\phi \Mpl}$ is the temperature of the onset of the axion oscillation.
Thus we have $m_\phi\gtrsim0.1{\rm meV}$ for the initial trapping at the onset of oscillation and if we take $\lambda\sim 1$ (i.e. $m_\Psi \sim 1$\,eV). This condition is similar to  the condition that the onset of oscillation should happen before the electroweak transition.

For $m_\phi \lesssim 0.1\,$eV, we need to consider the trapping at $T<100\,$GeV and cancellation between
the contributions, $\lambda_{H \Psi}v^2_{\rm EW}-m_\Psi^2$, 
to have the tiny dark Higgs mass in the low-energy. 
This cancellation allows us to have a larger portal coupling.
To study this case, let us take the effective field theory by integrating out the SM particles with mass larger than $T$. 
In particular, by integrating out the SM-like Higgs boson, we obtain the following dimension 5 interaction between the dark Higgs and the SM fermion $\psi$,
 \beq {\cal L}_{\rm eff}\supset -\theta_{\Psi^2 H}  \frac{m_\psi}{v_{\rm EW}} |\Psi|^2\bar{\psi} \psi.\eeq 
Here $\theta_{\Psi^2 H}$ is the effective coupling between  $\psi$ and $|\Psi|^2$, and $m_\psi$ is the mass of the fermion. 
In our case, we have the relation \beq \theta_{\Psi^2 H} \sim \frac{\lambda_{H\Psi} v_{\rm EW}}{m_h^2}\eeq, 
with $m_h$ being the SM-like Higgs boson mass $m_h\sim 125\,$GeV.
There is also a loop-induced dimension 6 gauge boson interaction. For simplicity let us neglect this contribution, keeping in mind that it becomes important
at some temperatures.
Then, the loop of the SM fermion, $\psi$, gives a thermal mass of \beq 
\Delta m(T_{\rm osc}<100\,{\rm GeV})
\sim \sqrt{\theta_{\Psi^2 H }T^2 \frac{m_\psi^2}{v_{\rm EW}}}\sim 1\,{\rm eV} \left(\frac{\lambda_{H\Psi}}{10^{-10}}\right)^{1/2} \frac{m_\psi}{0.1\,{\rm GeV}}\frac{T_{\rm osc}}{0.1\,{\rm GeV}} \eeq  to the dark Higgs field. 
By requiring the thermal mass to be larger than $m_\Psi\sim 1\,$eV at the onset of oscillation we obtain \beq   m_\phi\gtrsim 10^{-14}{\rm eV} \left(\frac{T_{\rm osc}}{0.1\,\rm GeV}\right)^2.\eeq 
We note that if we require $T_{\rm osc}\ll 0.1\, \rm GeV$, we need to consider electron contribution which requires $\lambda_{H\Psi}\gg 10^{-10}.$

For a larger coupling we may need to take account of  thermalization of the dark sector since thermally populated charged dark Higgs will generate a screening mass to the dark photon, and the tachyonic production is suppressed. 
The dark Higgs sector would be thermalized at around the electroweak phase transition. 
The thermalization is most efficient at $T\sim 100$GeV. By requiring the thermalization rate $\frac{\lambda_{H\Psi}^2}{4\pi}T$ to be smaller than the Hubble expansion rate, we obtain 
\beq
\lambda_{H\Psi}\lesssim 10^{-8}. 
\eeq 
When $\lambda_{H\Psi} \gg 10^{-8}$ the dark Higgs number can be still suppressed if the reheating temperature is significantly lower than 100\, GeV. 
Then a relatively larger $\lambda_{H\Psi}$ is compatible with our scenario. In this region, 
we need to be careful of the phenomenological and astrophysical bounds. Since $\Psi$ is light, the pair production of $\Psi \bar \Psi$ in the accelerator experiments and stars should be the dominant process. 
For instance, the pair emission rate of  $\Psi \bar\Psi$ in the SN1987A very roughly is $\epsilon_{\Psi \bar \Psi}\sim \frac{\lambda^2_{H \Psi} T^5}{ m_h^4}\sim  10^{29}{\rm erg \cdot g^{-1}\cdot s^{-1}}\lambda_{H\Psi}^2\left(\frac{T}{30\rm \, MeV}\right)^5$ from the dimensional argument, with $T\sim 30\,$MeV being the typical core temperature of the proto-neutron star. We have neglect some  extra  small factors for the phase space or SM coupling suppression in the rough estimate. 
This should satisfy the so-called Raffelt criterion $\epsilon_{\Psi \bar\Psi}\lesssim 10^{19} {\rm erg \cdot g^{-1}\cdot s^{-1}}$ \cite{Raffelt:1996wa}. Thus $\lambda_{H\Psi}\lesssim 10^{-5}$ is obtained. If $\lambda_{H\Psi}$ is much larger than this limit, the SN1987A bound might be alleviated since the $\Psi$ is trapped in the core. Such  large $\lambda_{H\Psi}$ in the trapped regime may also be constrained in the accelerator via, e.g., $K\to \pi +\Psi\bar\Psi.$

\bibliography{reference}

\end{document}